# Equilibrium condensation from chondritic porous IDP enriched vapor: Implications for Mercury and enstatite chondrite origins

D. S. Ebel [a, b,*], C. M. O'D. Alexander [c]

[a] Department of Earth and Planetary Science, American Museum of Natural History Central Park W. at 79th St., New York, NY 10024.

[b] Lamont-Doherty Earth Observatory, Columbia University, Palisades, NY 10964

[c] Dept. of Terrestrial Magnetism, Carnegie Institution of Washington, 5241 Broad Branch Rd., Washington, DC 20015 USA.

*debel@amnh.org

**Abstract:** The origin of Mercury's anomalous core and low FeO surface mineralogy are outstanding questions in planetary science. Mercury's composition may result from cosmochemical controls on the precursor solids that accreted to form Mercury. High temperatures and enrichment in solid condensates are likely conditions near the midplane of the inner solar protoplanetary disk. Silicate liquids similar to the liquids quenched in ferromagnesian chondrules are thermodynamically stable in oxygen-rich systems that are highly enriched in dust of CI-chondrite composition. In contrast, the solids surviving into the orbit of Mercury's accretion zone were probably similar to highly unequilibrated, anhydrous, interstellar organic- and presolar grain-bearing chondritic, porous interplanetary dust particles (C-IDPs). Chemical systems enriched in an assumed C-IDP composition dust produce condensates (solid + liquid assemblages in equilibrium with vapor) with super-chondritic atomic Fe/Si ratios at high temperatures, approaching 50% of that estimated for bulk Mercury. Sulfur behaves as a refractory element, but at lower temperatures, in these chemical systems. Stable minerals are FeO-poor, and include CaS and MgS, species found in enstatite chondrites. Disk gradients in volatile compositions of planetary and asteroidal precursors can explain Mercury's anomalous composition, as well as enstatite chondrite and aubrite parent body compositions. This model predicts high sulfur content, and very low FeO content of Mercury's surface rocks.

## 1. The problem of Mercury and the enstatite chondrites

The core of Mercury is at least 60% of that planet's mass, much higher than any of the other terrestrial planets. The origin of Mercury's anomalous density remains an outstanding problem in planetary science. Mercury's original mantle may have been partially stripped by impact(s) after planetary accretion (Wetherill, 1988; Benz et al., 1988, 2007). However, alternative hypotheses involving chemical and dynamical fractionation of the metal and silicates that accreted to form Mercury cannot be ruled out. Chemistry and dynamics in the radial zones from which the terrestrial planets accreted should vary with solar radius. Comprehensive modeling of such effects should account for phenomena including the similar oxygen isotopic signatures of the Earth-Moon system and enstatite chondrites (Clayton, 1993), the reduced oxidation state of the Martian mantle (Ghosal et al., 1998; Wadhwa, 2001, 2008), and the remnant chemical zoning of the asteroid belt (Gradie et al., 1989).

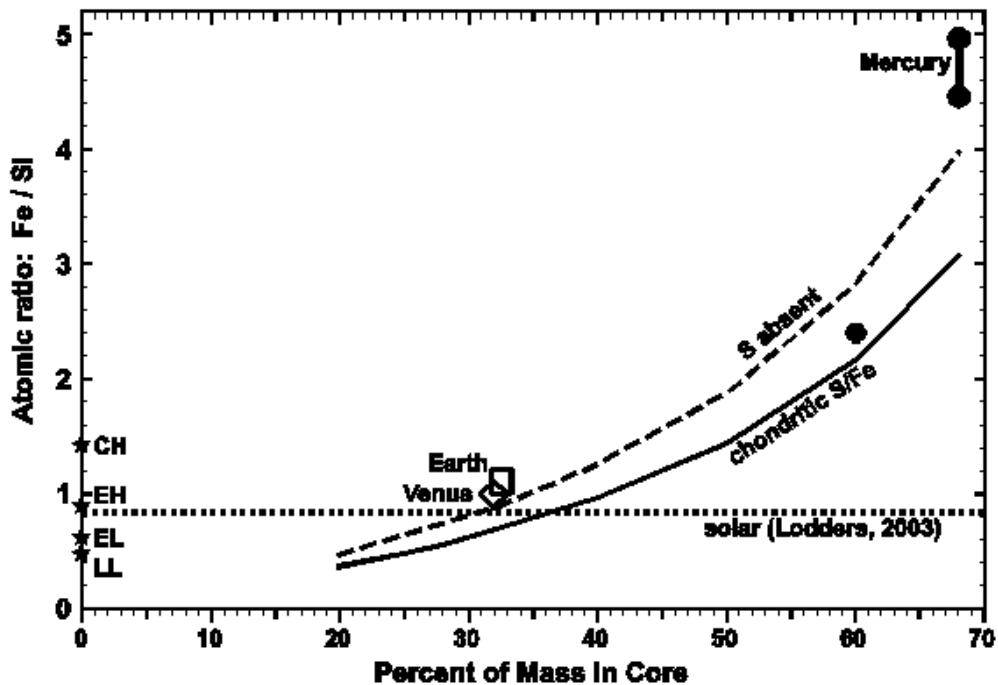

**Figure 1.** Estimated atomic Fe/Si ratios of meteorites, terrestrial planets, and calculated curve assuming chondritic bulk mantle and bulk core. Meteoritic values for EH, EL, LL and CH chondrites, and solar composition (Lodders, 2003) are plotted for reference. Single filled circle indicates minimum estimate for Mercury Fe/Si considered here.

Figure 1 illustrates Fe/Si (atomic) ratios calculated for Venus (Morgan and Anders, 1980), Earth (Kargel and Lewis, 1993), and Mercury (4.45, Goettel, 1988; 4.94, BVSP 1981, extreme collision model), relative to core mass fraction (Lodders and Fegley, 1998). Ratios for ordinary and enstatite chondrites (Wasson and Kallemeyn 1988) are also shown. The curves are calculated for each core fraction assuming separately chondritic (Lodders 2003) metallic core (Fe, Co, Ni, S) and oxide mantle (Mg, Si, Al, Ca, Na, Ti, O) compositions, and the same with no S present. These curves provide upper and lower bounds on Fe/Si in a fully reduced (no FeO in the mantle) Mercury. There is a high degree of uncertainty in both the core mass and the bulk composition of Mercury. Here, we adopt the minimum bound for the core mass of 60%, and a calculated Fe/Si of 2.4 for a core with S/Fe at 0.6×chondritic S/Fe (Fig. 1).

Potential Mercury precursor materials represented in the meteorite record vary in their metal/silicate ratios and degree of oxidation/reduction. The atomic Fe/Si ratio of the solar system is 0.838 (Lodders, 2003; Fe/Si = 0.900 according to Anders and Grevesse, 1989), and chondrites range from 0.492 (LL chondrites) to 0.873 (EH chondrites, Wasson and Kallemeyn, 1988), and up to 1.42 (CH chondrites, Lodders and Fegley 1998). This range demonstrates that there were processes in the early solar nebula that efficiently fractionated metal and silicates. The atomic Fe/Si ratio for Mercury is 2.18, assuming solar proportions of Mg, Si, Al, Ca, Na, Ti as oxides, and solar proportions of Fe, Ni, Co and S, all reduced and in the core, and assuming that 60% of Mercury's mass is in its core. If only 60% of S differentiates into the core, the planetary Fe/S ratio rises to ~2.4 (Fig. 1). Here, scenarios are explored which might produce solid assemblages with high Fe/Si ratios over a wide range of elevated temperatures that might be typical of the innermost solar system.

Similar scenarios also might explain both the high Fe/Si ratios of EH chondrites, and the reduced nature of all enstatite chondrites and the related magmatic aubrite achondrites. The EL enstatite chondrites (EC) on average have Fe/Si ratios of only 0.595 (Wasson and Kallemeyn, 1988), but all EC contain matrix minerals (e.g.-(Fe,Mg,Mn)S solid solution; and oldhamite, CaS) that probably formed in near thermodynamic equilibrium with a vapor phase. The aubrite meteorites also contain Ca-, and Mg-sulfides, and are considered by some to be partially differentiated, or melted, relatives of the enstatite chondrites (Keil, 1989). The aubrites, the EC, and Earth-Moon are all very similar in their oxygen isotopic compositions.

The abundances of volatile elements and compounds, particularly water ice and carbon, may have varied substantially in the inner disk during the accretion of Mercury. Here, we explore some effects of these abundance variations on the composition of condensates in equilibrium with vapor. The protoplanetary disk contained an isopleth of total pressure ($P^{tot}$) - temperature (T) conditions at which water sublimed or condensed as ice (e.g., Kornet et al., 2004; Podolak and Zucker, 2004; Lecar et al., 2006). This 3D 'snow line' moved radially and azimuthally as the disk evolved but is variously placed at 3 to 5 AU in the disk midplane during formation of the asteroid belt (Hueso and Guillot 2003). Water can be expected to condense into ices beyond the snow line, and sublime to vapor sunward of the snow line, producing a 'cold finger' effect (Stevenson and Lunine, 1988) that may be responsible for the extreme hydrogen depletion of the terrestrial planets and meteorites (Albaréde, 2009).

Particularly at late disk times, the midplane is expected to have been enriched in dust (condensable material), relative to a system with a solar composition (Cuzzi and Zahnle, 2004; Cuzzi and Weidenschilling, 2005). Within a few AU of the Sun, the midplane would have experienced temperatures and densities that were sufficient to chemically and isotopically equilibrate solids, liquids, and vapor on short timescales (Cassen, 2001). Here, we examine the effect of enrichment of the inner solar system by a carbon-rich, water/oxygen-poor dust produced by sublimation of ice as the dust migrated inward of the snow line. This dust would have been similar to unequilibrated, anhydrous, porous interplanetary dust particles (C-IDPs). Preliminary expositions of this idea were presented by Alexander (2002), and Ebel and Alexander (2002, 2005).

## 2. Approach to the problem

Ebel and Grossman (2000) explored the thermodynamic stability of FeO in silicates as a function of P, T, and enrichment in a dust of CI chondrite composition (~Orgueil, Anders and Grevesse, 1989). They found that enrichments in this dust would have stabilized ferromagnesian chondrule liquids. We use the same calculation methods here.

The CI chondrites are the most volatile-rich primitive meteorites, containing substantial bound water in hydrous silicates (18-22 wt%, Weisberg et al., 2006). They also contain oxidized phases, such as $Fe_3O_4$, and carbonate and sulfate minerals. These hydrous silicates and oxidized phases are thought to be the products of alteration by water of a more reduced, anhydrous primordial dust in the CI parent body. Studies of extraterrestrial materials suggest that good analogs for the primordial dust that was present in the early solar protoplanetary disk are the C-IDPs (Messenger and Walker, 1997; Messenger et al., 2003). The C-IDPs likely represent cometary ejecta and debris from C-rich asteroidal bodies, and are highly unequilibrated, anhydrous, contain abundant Fe-metal and C and have higher abundances of presolar materials than any chondrite (Schramm et al., 1989; Thomas et al., 1993; Huss, 1988; Huss et al., 2005; Busemann et al., 2009). The water that reacted with the dust in CI chondrites, and in other chondrites, was presumably accreted as ice, with chondritic matrix. Yet it can be expected that ices would sublimate from primitive, reduced dust grains as they migrated sunward past the snow line. Indeed, the enstatite chondrites are anhydrous, but contain C and presolar materials in their matrices in approximately the same proportions as in CI chondrites (Alexander et al., 1998, 2007; Huss, 1990).

Chemical systems of solar composition but with increased carbon (high C/O ratio) stabilize Ca- and Mg-sulfide minerals (Larimer, 1968, 1975; Larimer and Bartholomay, 179; Lodders and Fegley, 1993; Sharp and Wasserburg, 1995; Ebel, 2006). Wood and Hashimoto (1993) first examined the consequences of increasing C content in dust-enriched systems but did not explain how such chemical systems might occur. The sublimation of ice and subsequent concentration of the remaining relatively carbon-rich and oxygen-poor dust is a natural mechanism for achieving high C/O ratios in the innermost solar system.

### 2.1. Initial compositions

The starting compositions for the condensation calculations use a C-IDP dust analog. This is an H-, N-, F-, Cl-free dust of Orgueil (CI) composition (Anders and Grevesse, 1989), retaining all C as elemental (~graphitic) carbon, all S as FeS, Co and Ni as metal, and only enough O to make oxides of the remaining Fe and of all Si, Mg, Ca, Al, Na, P, K, Ti, Cr, Mn. This C-IDP dust is almost identical to the CI dust considered by Ebel and Grossman (2000), with most elements at chondritic ratios (e.g.-Mg/Si=1.074). The major difference is that this C-IDP dust contains less O than CI dust (Table 1), but remains more oxidizing than true, FeO-poor C-IDPs (Thomas et al. 1993). The present calculation also assumes the solar abundances of Lodders (2003, Table 1, col 3), which incorporates

C and O abundances (Allende Prieto et al., 2002) that are lower than those of Anders and Grevesse (1989). The analog dust was added to complementary solar composition vapor (enrichment of 1 = solar) in various proportions to make the bulk compositions.

**Table 1:** Bulk compositions (atoms, normalized to Si = $10^6$) considered in calculations. Solar compositions of Anders and Grevesse (1989) and Lodders (2003) are presented for comparison.

|    | solar (AG89) | solar (L03) | 1000x CI | 1000x C-idp |
|----|---|---|---|---|
| H  | 2.79E+10 | 2.88E+10 | 3.32E+07 | 2.88E+07 |
| He | 2.72E+09 | 2.29E+09 | 2.72E+06 | 2.29E+06 |
| C  | 1.01E+07 | 7.08E+06 | 7.65E+05 | 7.62E+05 |
| N  | 3.13E+06 | 1.95E+06 | 6.28E+04 | 1.95E+03 |
| O  | 2.38E+07 | 1.41E+07 | 7.65E+06 | 3.73E+06 |
| F  | 8.43E+02 | 8.41E+02 | 8.43E+02 | 8.41E-01 |
| Ne | 3.44E+06 | 2.15E+06 | 3.44E+03 | 2.15E+03 |
| Na | 5.74E+04 | 5.75E+04 | 5.74E+04 | 5.61E+04 |
| Mg | 1.07E+06 | 1.02E+06 | 1.07E+06 | 1.03E+06 |
| Al | 8.49E+04 | 8.41E+04 | 8.49E+04 | 8.48E+04 |
| **Si** | **1.00E+06** | **1.00E+06** | **1.00E+06** | **1.00E+06** |
| P  | 1.04E+04 | 8.37E+03 | 1.04E+04 | 1.00E+04 |
| S  | 5.15E+05 | 4.45E+05 | 5.15E+05 | 4.31E+05 |
| Cl | 5.24E+03 | 5.24E+03 | 5.24E+03 | 5.24E+00 |
| Ar | 1.01E+05 | 1.03E+05 | 1.01E+02 | 1.03E+02 |
| K  | 3.77E+03 | 3.69E+03 | 3.77E+03 | 3.81E+03 |
| Ca | 6.11E+04 | 6.29E+04 | 6.11E+04 | 5.92E+04 |
| Ti | 2.40E+03 | 2.42E+03 | 2.40E+03 | 2.40E+03 |
| Cr | 1.35E+04 | 1.29E+04 | 1.35E+04 | 1.35E+04 |
| Mn | 9.55E+03 | 9.17E+03 | 9.55E+03 | 9.49E+03 |
| Fe | 9.00E+05 | 8.38E+05 | 9.00E+05 | 8.72E+05 |
| Co | 2.25E+03 | 2.32E+03 | 2.25E+03 | 2.26E+03 |
| Ni | 4.93E+04 | 4.78E+04 | 4.93E+04 | 4.93E+04 |

*2.2. Calculations*

Thermodynamic equilibria between fully speciated vapor, silicate liquid, and solid minerals were calculated at 5 or 10° steps at fixed $P^{tot} = 10^{-4}$ bar for each bulk composition (Ebel et al., 2000). The results for CI-composition dust are taken from Ebel and Grossman (2000; cf., Ebel, 2006). Only conservative dust enrichment factors of up to 1000 are investigated here. However, recent work suggests that dust enrichment may have been much higher during chondrule formation (Alexander et al., 2008). We restrict the range of FeO and C contents of the C-IDP dust, total pressures, and dust enrichment factors considered here in order to establish with concision the plausibility of the hypothesis.

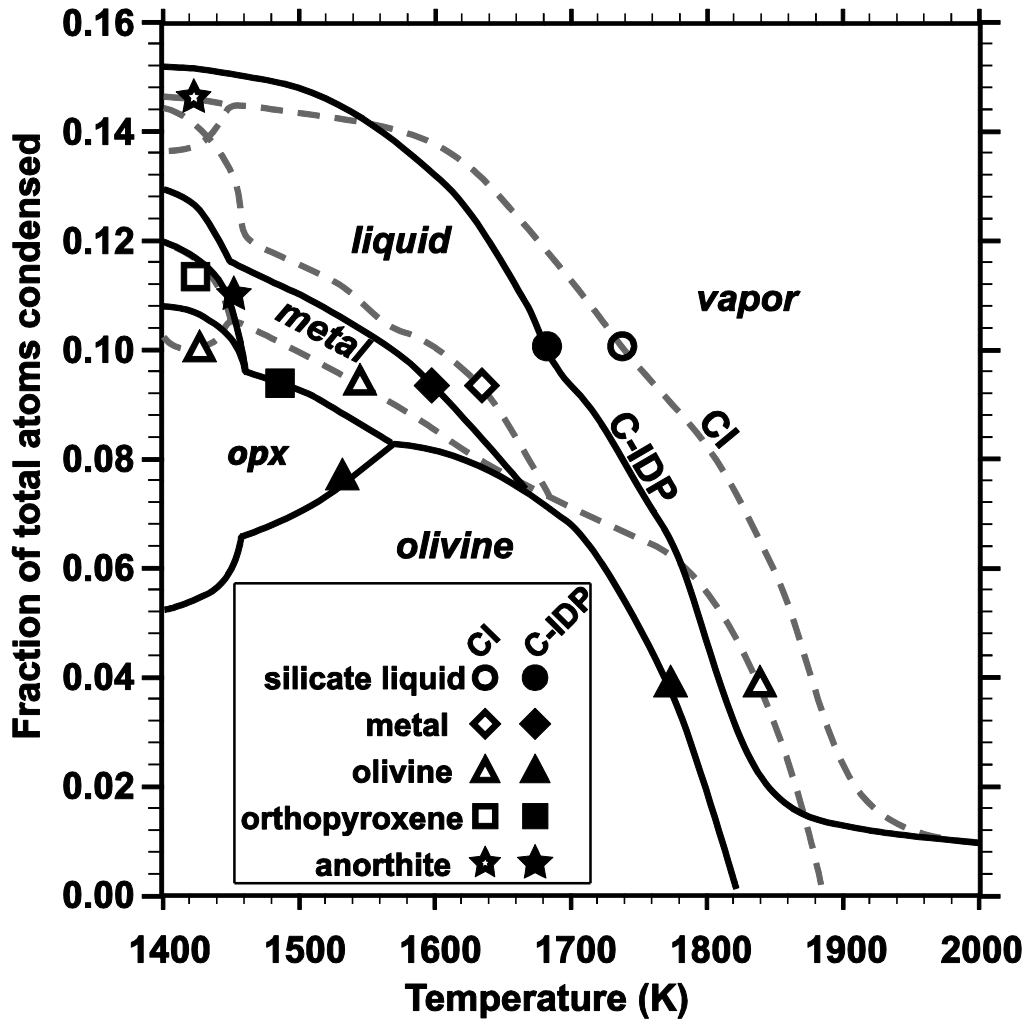

**Figure 2.** Cumulative fraction of atoms condensing in major phases from vapors of solar composition enriched 1000x in CI dust (dashed gray lines, open symbols) and C-IDP dust (solid lines, filled symbols), at $P^{tot} = 10^{-4}$ bar. Phases are liquid (circles), olivine (triangles), metal alloy (diamond), orthopyroxene (squares), and feldspar (stars).

## 3. Results

*3.1. Iron enrichment of condensates*

A dust enrichment of 1000x CI (atomic), where CI is 'solar', corresponds to a dust/gas *mass* ratio of only ~6.7, a not unreasonable expectation for the protoplanetary disk midplane (Ebel, 2006). Enrichment to the same degree in C-IDP dust (~5.1x by mass) leads to very different midplane chemistry. Figure 2 illustrates the balance of atoms between coexisting phases, and Fig. 3 shows the atomic Fe/Si ratio in the bulk condensed fraction with decreasing temperature. At 1000x enrichment in C-IDP dust, the Fe/Si ratio in the condensed portion remains above 1 during cooling from 1595 K to 1400 K. In systems that are enriched 1000x in chondritic (CI) dust the Fe/Si ratio does not rise above the solar bulk ratio of the total system, but the ratio of Fe in metal and sulfide, to Si in silicates, is much lower due to the oxidizing nature of chondritic dust. By comparison, in a vapor of solar composition (Anders and Grevesse, 1989), which is more reducing than a system enriched in CI-type dust, only a narrow range of temperatures exists where there is an elevated Fe/Si ratio in the condensed fraction, which only makes up $1.6 \times 10^{-5}$ of the total chemical system at 1400 K. Hence, it is not illustrated in Fig. 2.

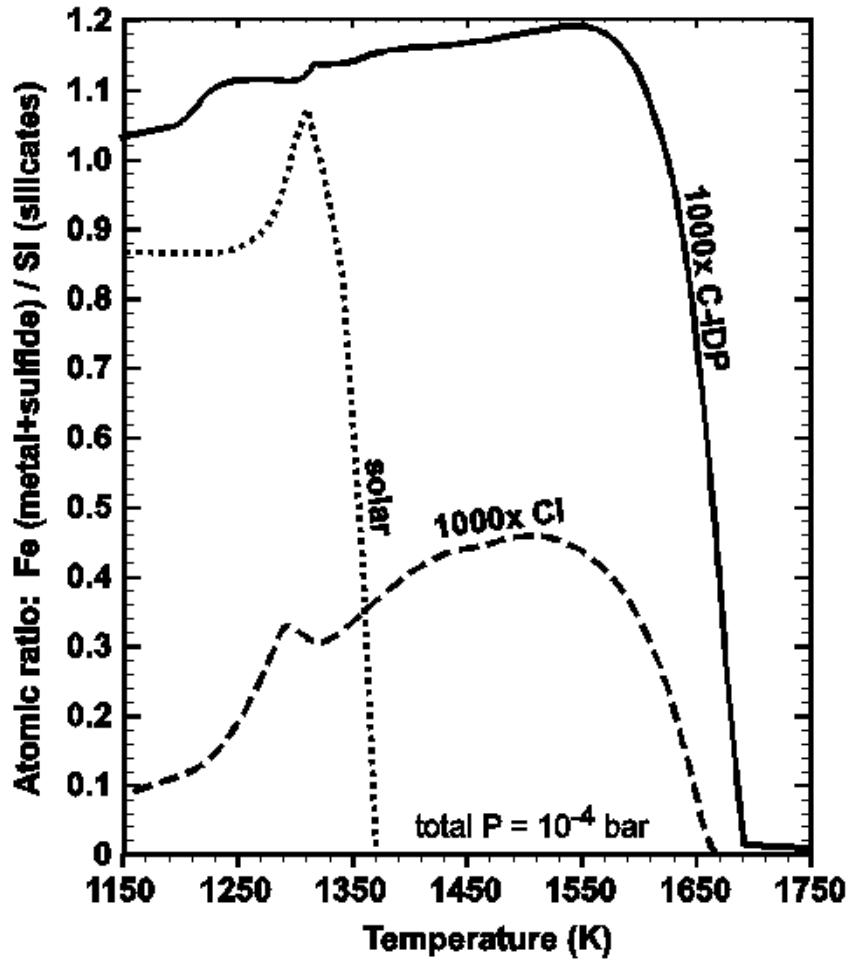

**Figure 3.** Atomic ratio Fe (in metal and sulfide)/Si (in silicates) in bulk condensates, for the same bulk compositions as in Fig. 2 as well as a vapor of solar composition, at $P^{tot}$ = $10^{-4}$ bar. Condensation of 1000x CI enriched vapor (dashed line) is calculated using the solar composition of Anders and Grevesse (1989) with bulk Fe/Si = 0.90 (Ebel and Grossman, 2000).

*3.2. Condensation of reduced mineral assemblages*

In C-IDP-enriched systems, condensing Mg-rich olivine draws O from the vapor, sharply decreasing $f(O_2)$ as the system cools. This is illustrated in Fig. 4, which also illustrates $f(O_2)$ in a condensing vapor of solar composition (Ebel and Grossman, 2000), using the solar composition of Anders and Grevesse (1989). As a result, the FeO content of the condensed silicates *decreases* with decreasing temperature in C-IDP enriched systems, in contrast to CI-dust enriched systems (Ebel and Grossman, 2000, their Fig. 8). This behavior is also evident in Fig. 3, where Fe as metal and sulfide increases in C-IDP enriched systems, but decreases in CI-dust enriched systems as FeO enters silicates with decreasing temperature. Systems become so reduced that at $P^{tot}=10^{-4}$ CaS becomes stable at 1340 K at 100x C-IDP dust, and at 1365 K at 1000x C-IDP dust; and MgS at 1195 K and 1230 K for the same C-IDP dust enrichment factors (Fig. 5). At 1000x C-IDP dust enrichment, the proportion of the total Ca (atoms) present as CaS increases from ~20% at 1360 K to 90% at 1300 K, while the fraction of Mg as MgS rises from 1.5% at 1230 K to 10% at 1200 K.

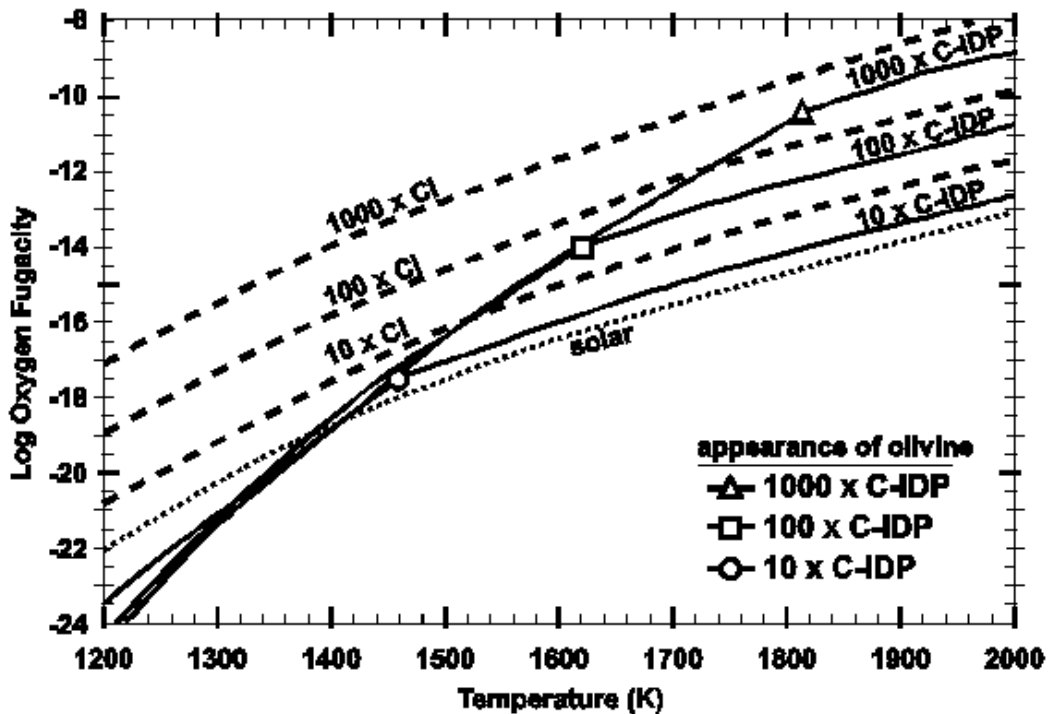

**Figure 4.** Oxygen fugacities in cooling systems enriched in CI composition dust (dashed lines), and C-IDP dust (solid lines), at $P^{tot} = 10^{-4}$ bar. At high temperatures (vapor only), the $f(O_2)$ differs by one log unit for constant dust enrichment but two different dust compositions. Open symbols mark the appearance of olivine in C-IDP enriched systems, where $f(O_2)$ dips steeply upon cooling. Results for a vapor of solar composition (Anders and Grevessse, 1989) are also shown.

　　In the C-IDP dust enriched system, with declining $f(O_2)$, FeO becomes *less* stable as a component of the condensates, and orthopyroxene dominates over olivine in the condensate assemblage. The calculated stability fields in Fig. 5 are quite similar in general, but shifted down in temperature, with increasing enrichment in less oxidizing C-IDP dust. The only sulfide that appears in the CI dust-enriched case is FeS, and then only at high enrichments in dust.
　　The reducing nature of the C-IDP-enriched composition is apparent in the details of mineral composition. At 1000x CI dust enrichment (Ebel and Grossman 2000),

C/O=0.100, Si/O=0.131. At 1000x C-IDP dust enrichment, C/O=0.204, Si/O=0.268. At 1000x dust, at $P^{tot}=10^{-4}$ bar, >99% of Si speciates as $SiO_{(g)}$ above 1900 K. At 1000x CI dust, at 1400 K, Si is 7% in silicate liquid, 70% in olivine, 16% in orthopyroxene (opx), the remainder in anorthite and Ca-rich pyroxene, with the FeO component of minerals $X_{fa}$=0.32 in olivine, $X_{fs}$=0.24 in opx. For 1000x C-IDP dust at this T and P ($10^{-4}$ bar), Si is ~7% in liquid, but only 28% in olivine, 44% in opx, 7% in anorthite, and ~13% remaining in the vapor. At 1400 K, $X_{fa}$ and $X_{fs}$ are <0.001, and over 90% of the 13% Si in the vapor is speciated as $SiS_{(g)}$, the remainder as $SiO_{(g)}$. With decreasing T, olivine is converted to opx by reaction with gaseous Si. Olivine disappears at 1200 K, but ~3% of Si remains in the vapor as $SiS_{(g)}$ at 1200 K.

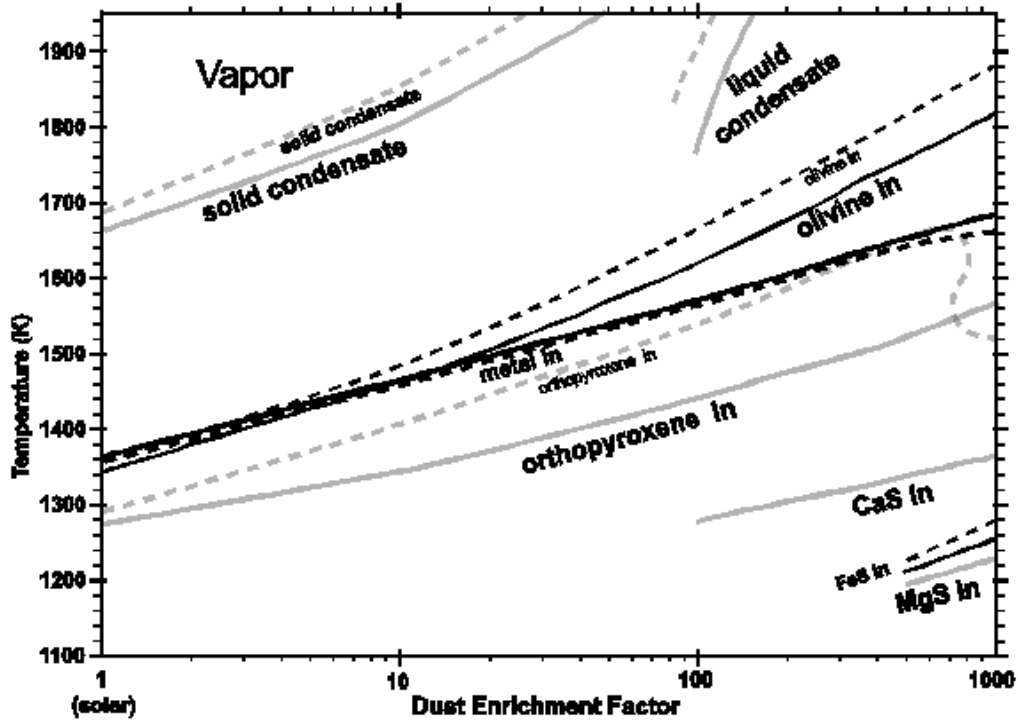

**Figure 5.** Appearance temperatures of major condensed phases with increasing dust enrichment, for CI (dashed lines) and C-IDP (solid lines) dust compositions, at $P^{tot} = 10^{-4}$ bar. Data are interpolated between results at 1, 10, 100, 500 and 1000x enrichment. The liquid stability fields have not been calculated in detail, but solidi are all at ~1380 K. Grey lines indicate appearance temperatures for first solid condensate, silicate liquid, orthopyroxene, CaS and MgS. Darkest lines indicate olivine, metal, and FeS.

## 4. Discussion

Silicate liquids and solids that are similar in composition to ferromagnesian chondrules are thermodynamically stable in O-rich systems that are highly enriched in dust of CI-chondrite composition (Ebel and Grossman, 2000; Ebel, 2006). Chemical systems enriched in a dust that is similar in composition to C-IDPs are more reducing. These systems produce condensates (solid + liquid assemblages in equilibrium with vapor) with superchondritic atomic Fe/Si ratios that approach 50% of the estimated ratio for bulk Mercury of 2.4. Stable minerals are FeO-poor, and include CaS and MgS, species found in enstatite chondrites.

This calculation is very conservative in assuming the composition of C-IDP dust. Rietmeijer (2002) notes that measured C contents of IDPs represent lower limits due to atmospheric entry effects. Recalculation of all 19 compositions reported by Thomas et al. (1993) yields atomic ratios Fe/Si = 1.01 ± 1.77, C/O = 0.53 ± 0.28, S/O = 0.10 ± 0.06, compared to the solar ratios 0.84, 0.50, and 0.03, respectively (Lodders, 2003), and the ratios in our assumed C-IDP dust, 0.87, 0.20, and 0.12, respectively (Table 1). A system enriched in a dust like the average IDP of Thomas et al. (1993) would exhibit intrinsically higher Fe/Si in the resulting condensates.

The reason for the high Fe/Si ratios in high temperature condensates in the C-IDP enriched system is that significant Si remains in the vapor above 1200 K as the species $SiS_{(g)}$, after all Fe has condensed, and because negligible FeO condenses into silicates at any temperature. In oxidized systems (e.g.-solar, CI dust-enriched), after $H_2O$ and the very stable molecule $CO_{(g)}$ form, most of the excess O forms $SiO_{(g)}$. In the C-IDP enriched systems, $SiO_{(g)}$ is less stable relative to $CO_{(g)}$, with declining $f(O_2)$. Thus $CO_{(g)}$ starves $SiO_{(g)}$ of O, and the excess Si forms $SiS_{(g)}$. However, $SiS_{(g)}$ is more stable, relative to silicate solids and liquids, than $SiO_{(g)}$, so in the C-IDP-enriched systems, significant Si remains in the vapor phase to lower temperatures than in the CI-enriched systems.

*4.1. Implications for Mercury*

*4.1.1. Mantle FeO*

Mercury may have accreted from a radial annulus of solid materials that equilibrated with vapor in a chemical system highly (> 100x) enriched in a dust of C-IDP composition. This result is consistent with nearly all Fe in Mercury being in the core, and less than 2 wt% FeO in mantle silicates. Many remote observations support a very low FeO content in Mercury's surface rocks (Boynton et al., 2007), and perhaps a radial gradient in oxidation state in the solar nebula (Robinson and Taylor, 2001). The model presented here predicts that Mercury formed from highly reduced precursor material. This prediction will be testable by measurements of total surface Fe by the x-ray spectrometer on the MESSENGER spacecraft.

*4.1.2. Sulfur enrichment*

This work predicts that Mercury should not follow the systematic depletions in S apparent among asteroids, Mars, Venus and Earth (Taylor, 1988). Enrichment in C-IDP dust results in more refractory behavior of S than in a vapor of solar composition, because sulfides of Ca and Mg are stable phases in equilibrium with vapor at high temperatures in systems that are enriched in C-IDP dust. If the innermost planet accreted from a narrow radial annulus from condensates of this kind, then Mercury should have a higher bulk S abundance than Earth, relative to chondritic abundances. Similar arguments may apply to Io in a Jupiter subdisk.

Although CaS has a melting point well above the Fe-FeS eutectic (1263 K), it dissolves completely above ~1273 K in EC melts (McCoy et al., 1999). The behavior of such reduced melts is complex, but the Fe-deficient nature of aubrite meteorites, the immiscibility of Fe-rich and Fe-poor sulfide liquids, the high solubility of S in reduced silicate melts, and the crystallization of CaS from those melts at high T all suggest that a significant fraction of Mercury's S remained in the mantle (Fogel et al., 1996; McCoy et al., 1999). This prediction will be testable by measurements of S by the x-ray spectrometer on the MESSENGER spacecraft. A correlation between Ca and S might be expected, rather than between Ca and Al as predicted in oxidized systems, although the strength of the Ca-S correlation will be a function of the degree of reduction.

*4.1.3. Iron/silicate ratio*

　　Hypotheses for Mercury's high core/mantle ratio have been reviewed elsewhere (Taylor and Scott, 2003; Boynton et al., 2007). These include aerodynamic sorting (Weidenschilling, 1978), preferential vaporization of silicates and removal by solar wind (Fegley and Cameron, 1987), incomplete (fractional) condensation of silicates (Lewis, 1973), and mantle stripping by giant impact (Wetherill, 1988). The hypothesis tested here is a variant on condensation models. All of these hypotheses, and the present work, predict distinct chemical consequences for Mercury's mantle composition that should be testable from MESSENGER observations. Burbine et al. (2002) predicted an enstatite basalt composition based on Mercury's spectral similarity to highly reduced aubrite meteorites. This work provides a rationale for how such a planetary composition might occur.

　　There are systematic depletions in volatile elements among meteorite parent bodies (asteroids) and the terrestrial planets (Dreibus and Palme, 1996). This trend is seen in plots of element abundances versus the 50% condensation temperature (Lodders 2003) for a vapor of solar composition at $P^{tot}=10^{-4}$ bar (e.g., for silicate Earth, McDonough, 2003). Our results show that for a vapor enriched in C-IDP dust, Si (but not Mg) is more volatile, over a broader temperature range, than in a vapor of solar composition (Fig. 3), and Fe is less volatile. In a system enriched 1000x in CI dust, the 50% condensation temperature ($T_{50}$) of Si is 1825 K and decreases to $T_{50}$ ~1712 K for 1000x C-IDP dust. For Fe, at 1000x CI dust, $T_{50}$ = 1455 K, but for 1000x C-IDP dust, $T_{50}$ = 1645 K.

　　Although conditions at the midplane of the inner solar nebula are poorly constrained, astrophysical models suggest peak temperatures above 1000 K to 2 AU for a minimum (0.02 solar mass) disk, and above 1500 K to 4 AU in a massive (0.13 solar mass) disk (Boss 1996a,b). Bell et al. (2000) conclude that the inner 1 AU exceeded 1000 K during the solar nebula's first few $10^5$ years. The timescale of planetary embryo formation is unknown, but embryos of Mercury's size may form on this timescale (Bottke et al. 2005). Indeed, Mercury may represent a remnant embryo from this nebula stage (Bottke et al. 2005). Models also indicate that nebular midplane pressures are highest inside 1 AU. Equilibrium condensation calculations show that at $P^{tot} > 10^{-4}$ bar, metal is stable at higher T (100 K at $10^{-1}$ bar) than olivine in a vapor of solar composition (Ebel, 2006, plate 1). This effect would enhance the high-temperature fractionation predicted here for $P^{tot} = 10^{-4}$ bar.

　　A high Fe/Si bulk ratio in Mercury by the condensation scenario considered here requires accretion of solids into Mercury or its precursors at T > 1200 K. This fractional condensation would result in a mantle enriched in FeO-free olivine, since in these systems Si is lost to the vapor. Because S is incompletely condensed into CaS at these temperatures, fractional condensation would also decrease the S content of Mercury. If Mercury's surface is dominated by orthopyroxene rather than olivine, this would rule out a fractional condensation origin for the planet's high Fe/Si ratio.

*4.2. Implications for reduced meteorites*

In these analog C-IDP dust-enriched systems, reduced phases (CaS, MgS) become stable in equilibrium with orthopyroxene above 1000 K, at dust enrichments >300x at $P^{tot}=10^{-4}$ bar. Calcium-, Al-rich oxides and silicates in these systems include the same minerals as calculated for CI dust-enriched systems (Ebel and Grossman, 2000), and FeO-free olivine, pyroxenes and feldspar are all calculated to condense at higher temperatures than CaS and MgS. Increased elemental C in C-IDP dust-enriched systems stabilizes mineral assemblages observed in EC and aubrites, relative to the oxidized assemblages seen in the ordinary and carbonaceous chondrites. However, unlike planets, the unequilibrated EC contain C in their matrix, along with presolar and other grains that would not have survived extreme heating. This mineral dust must have been accreted at low temperatures, precluding accretion in a region with very high ambient temperatures (Alexander et al., 1998, 2007; Huss, 1990).

If the dust composition (i.e., ice-containing or not) at the midplane was a function of the position of the snow line, then these results indicate that EC and aubrite parent bodies formed inside the snow line, where dust included C but not ices, particularly not water ice. The similar O isotopic signatures of the Earth-Moon system and the EC and aubrites suggest that they all formed from a similar oxygen isotopic reservoir. However, the strong depletion of the Earth in S suggests that Earth cannot have formed from material that experienced conditions so reducing that CaS, MgS and FeS were the major S-bearing phases, as is the case in EC and aubrite meteorites.

## 5. Conclusions

Mercury may have accreted from a narrow, hot, inner annulus that was highly enriched, relative to a vapor of solar composition, in a C-rich, $H_2O$-poor dust similar in composition to anhydrous chondritic interplanetary dust particles (C-IDPs). Calculations based on this scenario allow the prediction that Mercury is less depleted in S than the other terrestrial planets, and to have very little Fe (as oxide) in its mantle. In C-IDP dust enriched systems, Si becomes more volatile, and FeO is absent from condensed silicate minerals. Mercury's mantle may be enriched in S due to the high solubility of S in highly reduced silicate melts formed from CaS- and MgS-bearing mineral assemblages. The mineralogy and redox state of enstatite chondrites, and the Fe/Si ratios of EH chondrites, are consistent with their formation under similar conditions. A less conservative set of assumptions for C-IDP dust composition may partially explain Mercury's large core and high bulk metal/silicate ratio.

## Acknowledgements

This research has made use of NASA's Astrophysics Data System Bibliographic Services. Research was supported by the American Museum of Natural History, and National Aeronautics and Space Administration grants NNX10AI42G (DSE) and NNX09AF78G (CA).